\documentclass[11pt,a4paper]{article}

\def\AmSTeX{\leavevmode\hbox{$\mathcal A\kern-.2em\lower.376ex%
        \hbox{$\mathcal M$}\kern-.2em\mathcal S$-\TeX}}
\usepackage{graphicx}
\usepackage{amsmath}        
\usepackage{amssymb}
\usepackage{mathrsfs}       
\usepackage{bm}             
\usepackage[amsmath,thmmarks,hyperref]{ntheorem}
\newif\ifpdf \pdftrue
\ifx\pdfoutput\undefined \pdffalse \fi \ifx\pdfoutput\relax
\pdffalse \fi
\ifx\texonly\undefined\let\texonly\relax\fi
\ifx\endtexonly\undefined\let\endtexonly\relax\fi \texonly
  \let\htmlonly\iffalse
  \let\endhtmlonly\fi
\endtexonly

\htmlonly
  
\endhtmlonly
\texonly
  \usepackage{makeidx}

  \usepackage{multicol}
  

\endtexonly
\title{}
\author{\thanks{}}
\date{}
\begin{document}

\title{Estimating Form Factors of $B_s\rightarrow D_s^{(*)}$ and their\\
Applications to Semi-leptonic and Non-leptonic Decays}

\author{Xiang-Jun Chen$^a$,~~ Hui-feng Fu$^a$,~~ C. S. Kim$^b$\footnote{cskim@yonsei.ac.kr},
~~ Guo-Li Wang$^{a,c}$\footnote{gl\_wang@hit.edu.cn}\\
{\it \small  $^a$ Department of Physics, Harbin Institute of
Technology, Harbin, 150001, China} \\
{\it \small  $^b$ Department of Physics $\&$ IPAP, Yonsei
University, Seoul 120-749, South Korea}\\
{\it \small $^c$ PITT PACC, Department of Physics $\&$ Astronomy,
University of Pittsburgh, PA 15260, USA}}

\maketitle

\baselineskip=20pt
\begin{abstract}
\noindent $B_s^0\rightarrow D_s^{-}$ and
$B_s^0\rightarrow D_s^{*-}$ weak transition form factors are estimated
for the whole physical region with a method based on an
instantaneous approximated Mandelstam formulation of transition
matrix elements and the instantaneous Bethe-Salpeter equation.
We apply the estimated form factors
to branching ratios, CP asymmetries and polarization fractions of
non-leptonic decays within the factorization approximation.
And we study the non-factorizable effects and annihilation contributions
with the perturbative QCD approach. The
branching ratios of semi-leptonic $B_s^0\rightarrow
D_s^{(*)-}l^+\nu_l$ decays are also evaluated. We show that the
calculated decay rates agree well with the available experimental
data. The longitudinal polarization fraction of $B_s\rightarrow
D_s^*V(A)$ decays are $\sim0.8$ when $V(A)$ denotes a light meson,
and are $\sim0.5$ when $V(A)$ denotes a $D_q$ ($q=d,s$) meson.
\end{abstract}

\section{Introduction}

In the past few years, charmless non-leptonic $B_s$ decays have
been extensively studied~\cite{Charmless}, however,  the decays of
$B_s$ to charmed particles are relatively less studied. Therefore,
it is of urgent interest to put more attention on this topic.
Semi-leptonic $B_s^0\rightarrow D_s^{(*)-}l^+\nu_l$ decays and
non-leptonic ${B}^0_s\rightarrow D_s^{(*)-}X^+$ two body decays
(and their conjugated ones), where $X^+$ denotes a light meson or
a $D_q$ ($q=d,s$) meson, can reveal useful information about the
Cabibbo-Kobayashi-Maskawa (CKM) phases, the $B_s^0-\bar{B}_s^0$
mixing parameters~\cite{Aleksan,Fleischer1}, and the physics of CP
violations~\cite{Fleischer2}. Studies on $B_s$ decays to charmed
particles can be used to check the factorization
hypothesis~\cite{Xing} and to search for physics beyond the
Standard Model (SM)~\cite{Fleischer3,Fleischer4,Kim}.

For weak decays, the transition form factors play an important
role. Theoretically, to estimate the form factors of a relevant
process, one has to rely on some non-perturbative approaches such
as the Bethe-Salpeter (B-S) equation, quark models, QCD sum rules
(QCDSR) and lattice QCD. Turn to $B_s\rightarrow D_s^{(*)}$ weak
transitions, several works have been done: Early works, $e.g.$
\cite{Kramer,Bijnens}, usually relied on the famous
Bauer-Stech-Wirbel (BSW) model \cite{Bauer}. Authors of
\cite{Blasi,Azizi} adopted QCDSR for the calculation. In
\cite{GangLi}, the form factors are estimated within the covariant
light-front quark model (CLFQM). Authors of \cite{Li1} used the so
called light cone sum rules (LCSR) to investigate form factors at
large recoil, and heavy quark effective theory (HQET) to describe
them at small recoil region. Each of these methods sketches one or
another profile of non-perturbative QCD, and each has advantages
as well as shortcomings. So it is worthy to estimate the
$B_s\rightarrow D_s^{(*)}$ form factors in another method which is
based on the B-S equation \cite{Salpeter1} and the Mandelstam
formulation \cite{Mandelstam} of the transition matrix element. To
make predictions on non-leptonic decays, there is another task,
that is how to evaluate the decay amplitudes with form factors
available. It is well known that factorization approximation
(FA)~\cite{Bauer} has been extensively applied in non-leptonic
weak decays and has been justified to success in explaining the
branching ratios of several color-allowed $B_q$ decays~\cite{Ali}.
The works mentioned before all adopted the FA to evaluate
non-leptonic decay amplitudes. However, estimations based on the
FA still suffer uncertainties from the non-factorizable effects
and annihilation diagrams contributions, especially for the CP
asymmetries (CPAs). Thus approaches beyond the FA are in need.
Till now several approaches which can cover the non-factorizable
effects have been developed, such as the perturbative QCD (pQCD)
approach ~\cite{PQCD}, the QCD improved factorization (QCDF)
approach ~\cite{QCDF} and SCET approach ~\cite{SCET}. Studies on
$B_s$ decays into charmed particles with the perturbative QCD
approach have been carried out in ~\cite{Li2}. In this work, we
evaluate non-leptonic decay amplitudes under FA, as well as
estimate non-factorizable and annihilation contributions in the
pQCD approach. Besides these direct calculation in the FA or pQCD,
the authors in ~\cite{Ferrandes} used SU(3)$_F$ symmetry to
estimate the widths of a class of two-body $B_s$ decays with the
help of experimental data of corresponding $B$ decays. 
There have been some studies on semi-leptonic $B_s^0\rightarrow
D_s^{(*)-}l^+\nu_l$ decays with the approaches such as QCDSR~\cite{Azizi}, LCSR~\cite{Li1}, CLFQM~\cite{GangLi}
and constituent quark meson
model (CQM)~\cite{CQM}.

On the experiment side, the world averaged branching fractions of
some of the $B_s^0\rightarrow D_s^{(*)-}X^+$ decay modes are already
available~\cite{PDG}. Recently, Belle Collaboration has reported the
observations of $B_{s}^{0}\rightarrow D_s^{(*)-}D_s^{(*)+}$,
$D_s^{*-}\pi^+$ and $D_s^{(*)-}\rho^+$ decays and measurements of
their branching fractions~\cite{Belle}. Measurements of branching
ratio for $B_s^0\rightarrow D_s^{-}l^+\nu_l+$``anything" are given
to be $(7.9\pm2.4)\%$~\cite{PDG}. However, the exclusive
semi-leptonic decay rates for $B_s^0\rightarrow D_s^{(*)-}l^+\nu_l$
processes have not been measured yet. It is expected that in near
future more and more channels of $B_s$ decays will be precisely
measured experimentally.

In this paper, we estimate $B_s^0\rightarrow D_s^{(*)-}$ form
factors by calculating the corresponding transition matrix elements in the instantaneous approximated Mandelstam formulation with the wave functions obtained from the Salpeter equation. The Salpeter equation are derived from the B-S
equation in instantaneous approximation~\cite{Salpeter1}. The
benefits of this method are its firm theoretical basis and its
covering relativistic effects. It is well known that the B-S
equation is a relativistic two-body wave equation. With the form factors calculated,
we study the semi-leptonic $B_s^0\rightarrow
D_s^{(*)-}l^+\nu_l$ decays and the two body non-leptonic
$B_s^0\rightarrow D_s^{(*)-}X^+$ decays. For non-leptonic decays,
we estimate the branching ratios, CP asymmetries and polarization fractions of processes in the FA,
and we also estimate the non-factorizable and annihilation contributions with the pQCD approach for $B_s\rightarrow PP(PV,VP,VV)$ processes. Comparing our results with
different theoretical predictions and experiment data would enrich
our knowledge of $B_s$ weak decays to charmed mesons.

The paper is organized as follows: In Section 2, a brief review on
the Salpeter equation and the instantaneous approximated Mandelstam formulation
 of $B_s^0\rightarrow
D_s^{(*)-}$ transition matrix elements (from which the form factors are extracted) are presented. In Section 3,
we illustrate how we evaluate the decay amplitudes with the form factors calculated and how to evaluate the decay rates, CPAs and polarization fractions.
Section 4 is
devoted to numerical results and discussions.

\section{Form Factors of $B^0_s\rightarrow D_s^{(*)-}$ transitions}

Form factors are the crucial elements of decay amplitudes. In
order to estimate $B^0_s\rightarrow D_s^{(*)-}$ form factors,
first we use the improved Salpeter method illustrated in
~\cite{CHChang,W-C} to obtain the wave functions of $B_s$ and
$D_s^{(*)}$ mesons. In these literatures the authors solved the
full Salpter equations instead of only the positive energy part of
the equation. Now we give a brief review on this method. Under
instantaneous approximation, the well-known B-S equation
\begin{equation}(\not\!p_{1}-m_{1})\chi_{_P}(q)(\not\!p_{2}+m_{2})=i\int\frac{d^{4}k}{(2\pi)^{4}}V(P,k,q)\chi_{_P}(k)\end{equation}
can be deduced to be the full Salpeter equation, which equals to the following coupled equations \cite{Salpeter1}:
\begin{eqnarray}\label{eq2-n1}
&&(M-\omega_{1}-\omega_{2})\varphi_{_P}^{++}(q_{_{P_\perp}})=\Lambda_{1}^{+}(q_{_{P_\perp}})\eta(q_{_{P_\perp}})
\Lambda_{2}^{+}(q_{_{P_\perp}}),\notag\\
&&(M+\omega_{1}+\omega_{2})\varphi_{_P}^{--}(q_{_{P_\perp}})=-\Lambda_{1}^{-}(q_{_{P_\perp}})\eta(q_{_{P_\perp}})
\Lambda_{2}^{-}(q_{_{P_\perp}}),\notag\\
&&\varphi_{_P}^{+-}(q_{_{P_\perp}})=0,\ \
\varphi_{_P}^{-+}(q_{_{P_\perp}})=0.
\end{eqnarray}
Here $\chi_{_P}(q)$ is the B-S wave function of the relevant bound
state. $P$ is the four momentum of the state and $p_{1}$, $p_{2}$,
$m_{1}$, $m_{2}$ are the momenta and constituent masses of the quark
and anti-quark, respectively. $q$ is the relative momentum
$q=\alpha_2 p_1-\alpha_1 p_2$, where $\alpha_1 \equiv \frac{m_1}{
m_1 + m_2}$ and $\alpha_2\equiv\frac{ m_2}{ m_1 + m_2}$. $V(P,k,q)$
is the interaction kernel which can be written as
$V(k_{_{P_\perp}},q_{_{P_\perp}})$ under instantaneous
approximation. In our notations, $q_{_P}$ always denotes
$\frac{q\cdot P}{\sqrt{P^2}}$ and $q_{_{P_\perp}}=q-\frac{q\cdot
P}{P^2}P$. The definitions of $\eta(q_{_{P_\perp}}^{\mu})$,
$\Lambda_{1(2)}^{\pm}$ and $\omega_{1(2)}$ are gathered as follows:
\begin{eqnarray}
&&\eta(q_{_{P_\perp}}^{\mu})\equiv \int
\frac{\mathrm{d}k_{_{P_\perp}}^{3}}{(2\pi)^{3}}V(k_{_{P_\perp}}^{\mu},q_{_{P_\perp}}^{\mu})\varphi_{_P}(k_{_{P_\perp}}^{\mu}),~~~~
\varphi_{_P}(q_{_{P_\perp}}^{\mu})\equiv i\int
\frac{\mathrm{d}q_{_P}}{2\pi}\chi _{_P}(q),\notag\\
&&\Lambda_{1}^{\pm}=
\frac{1}{2\omega_{1}}[\frac{\not\!{P}}{M}\omega_{1}\pm(m_{1}+\not\!q_{_{P_\perp}})],~~~~\Lambda_{2}^{\pm}=
\frac{1}{2\omega_{2}}[\frac{\not\!{P}}{M}\omega_{2}\mp(m_{2}+\not\!q_{_{P_\perp}})],\notag\\
&&\omega_{1}=\sqrt{m_{1}^{2}-q_{_{P_\perp}}^{2}},~~~~~~~~\omega_{2}=\sqrt{m_{2}^{2}-q_{_{P_\perp}}^{2}}.
\end{eqnarray}
$\Lambda_{1(2)}^{\pm}$ satisfy
$\Lambda_{1(2)}^++\Lambda_{1(2)}^-=\frac{\not\!{P}}{M}$,
$\Lambda_{1(2)}^{\pm}\frac{\not\!{P}}{M}\Lambda_{1(2)}^{\pm}=\Lambda_{1(2)}^{\pm}$
and $\Lambda_{1(2)}^{\pm}\frac{\not\!{P}}{M}\Lambda_{1(2)}^{\mp}=0$.
With these $\Lambda^{\pm}$, the wave function $\varphi$ can be
decomposed into positive and negative projected wave functions
\begin{equation}\varphi_{_P}^{\pm\pm}(q_{_{P_\perp}})\equiv\Lambda_{1}^{\pm}(q_{_{P_\perp}})
\frac{\not\!{P}}{M}\varphi_{_P}(q_{_{P_\perp}})\frac{\not\!{P}}{M}\Lambda_{2}^{\pm}(q_{_{P_\perp}}).\end{equation}
In deriving the coupled equations (\ref{eq2-n1}), the decomposition
of the Feynman propagator
\begin{equation}\label{eq2-ad1}
S_j(p_j)=\left\{\frac{\Lambda_{j}^{+}(q_{_{P_{\perp}}})}{(-1)^{j+1}q_{_P}+\alpha_j
M-\omega_{j}+i\epsilon}
+\frac{\Lambda_{j}^{-}(q_{_{P_{\perp}}})}{(-1)^{j+1}q_{_P}+\alpha_j
M+\omega_j-i\epsilon}\right\},
\end{equation}
where $j=1$ for quark and $j=2$ for anti-quark has been used.

The wave functions relevant to $B_s\rightarrow D_s^{(*)}$ transition
have quantum numbers $J^P=0^-$ (for $\bar{B}_s^{0},B_s^0$ and
$D_s^{\pm}$) and $1^-$(for $D_s^{*\pm}$) and are written
as~\cite{CHChang,Wang}
\begin{eqnarray}
\varphi_{0^{-(+)}}(q_{_{P_\perp}})&=&M\left[\frac{\not\!P}{M}a_1(q_{_{P_\perp}})+{a}_{2}(q_{_{P_\perp}})
+\frac{\not\!q_{_{P_\perp}}}{M}{a}_{3}(q_{_{P_\perp}})+\frac{\not\!{P}\not\!q_{_{P_\perp}}}{M^2}{a}_{4}(q_{_{P_\perp}})\right]{\gamma}_5,
\\
{\varphi}^{\lambda}_{1^{-(-)}}(q_{_{P_\perp}})&=&(q_{_{P_\perp}}\cdot\epsilon^{\lambda})\left[b_1(q_{_{P_\perp}})+\frac{\not\!{P}}{M}b_2({q_{_{P_\perp}}})
+\frac{\not\!q_{_{P_\perp}}}{M}b_3({q_{_{P_\perp}}})+\frac{\not\!{P}\not\!{q_{_{P_\perp}}}}{M^2}b_4({q_{_{P_\perp}}})\right]+M\not\!{\epsilon}^{\lambda}b_5({q_{_{P_\perp}}})\notag\\
&&+\not\!{\epsilon}^{\lambda}\not\!{P}b_6(q_{_{P_\perp}})
+(\not\!{q_{_{P_\perp}}}\not\!{\epsilon}^{\lambda}-q_{_{P_\perp}}\cdot{\epsilon}^{\lambda})b_7({q_{_{P_\perp}}})
+\frac{1}{M}(\not\!{P}\not\!{\epsilon}^{\lambda}\not\!{q_{_{P_\perp}}}-\not\!{P}q_{_{P_\perp}}\cdot{\epsilon}^{\lambda})b_8({q_{_{P_\perp}}}),\notag
\end{eqnarray}
where $a_i(q_{_{P_\perp}})$ and $b_i(q_{_{P_\perp}})$ are wave
functions of $q_{_{P_\perp}}^2$; $M$ is the mass of corresponding
bound state; $\epsilon^\lambda$ is the polarization vector for
$J^P=1^{-}$ state. In numerical calculation, Cornell potential is
chosen as the kernel, and the explicit formulation is (in the rest
frame):
\begin{eqnarray}\label{eq2-ad7}
&\displaystyle V(\vec{q})=V_s(\vec{q})+V_v(\vec{q})\gamma^0\otimes\gamma_0,\notag\\
&\displaystyle
V_s(\vec{q})=-(\frac{\lambda}{\alpha}+V_0)\delta^3(\vec{q})
+\frac{\lambda}{\pi^2}\frac{1}{(\vec{q}^2+\alpha^2)^2},\notag\\
&\displaystyle
V_v(\vec{q})=-\frac{2}{3\pi^2}\frac{\alpha_s(\vec{q})}{(\vec{q}^2+\alpha^2)},
\end{eqnarray}
where the QCD running coupling constant
$\alpha_s(\vec{q})=\frac{12\pi}{33-2N_f}\frac{1}{\mathrm{log}(a+\vec{q}^2/\Lambda_{\mathrm{QCD}}^2)}$;
the constants $\lambda,\ \alpha,\ a,\ V_0$ and
$\Lambda_{\mathrm{QCD}}$ are the parameters characterizing the
potential, which are fixed by fitting the experimental mass
spectra. The parameters used in this work are $m_b=4.96$~GeV,
$m_c=1.62$~GeV, $m_s=0.5$~GeV, $\lambda=0.21$ GeV$^2$,
$\alpha=0.06$ GeV, $a=e=2.7183$, $\Lambda_{\mathrm{QCD}}=0.27$ GeV
and for $0^-$ state, $V_0=-0.432$ GeV $(c\bar{s})$, $-0.212$ GeV
$(b\bar{s})$, for $1^-$ state $V_0=-0.212$ GeV. With these
parameters, the wave functions of interested mesons can be
obtained by solving the coupled equations (\ref{eq2-n1}). The
details of how to solve the equation (\ref{eq2-n1}) could be found
in \cite{W-C}.

We now turn to evaluate weak transition form factors. The starting
point is the Mandelstam formulation of transition matrix elements.
Since the relevant wave functions are obtained from the B-S
equation with instantaneous kernel, instantaneous approximation
should be applied to the Mandelstam formulation which has been
carried out in details in \cite{Chang,Wang2,ZHWang}. We now follow
\cite{Chang} to sketch the derivation of the instantaneous
Mandelstam formulation as follows. According to Mandelstam
formalism, the transition matrix element between two bound states
induced by a current $\Gamma^\mu$, $e.g.$ $\gamma^\mu$,
$\gamma^\mu\gamma_5$, is written as
\begin{equation}\label{eq2-6}
\langle f(P_f) |(\bar{q}_1\Gamma^\mu
q_2)|i(P_i)\rangle=\int\frac{d^4q_i}{(2\pi)^4}\frac{d^4q_f}{(2\pi)^4}\mathrm{Tr}\Big[\bar{\chi}_{_{P_f}}^{f}(q_f)
\Gamma^\mu\chi_{_{P_i}}^{i}(q_i)iS_2^{-1}(p_{2i})\Big](2\pi)^4\delta^4(p_{2i}-p_{2f}),
\end{equation}
where $\bar{q}_1$ and $q_2$ are the relevant quark fields
operators. Here and hereafter in this section the superscript or
subscript $i$ and $f$ denote the quantities of the initial state
and the final state, respectively, in the transition. Using the
instantaneous B-S equation and decomposing the propagators into
positive and negative parts (see equation (\ref{eq2-ad1})),
equation (\ref{eq2-6}) can be deduced to
\begin{eqnarray}\label{eq2-ad2}
&&\langle f(P_f) |(\bar{q}_1\Gamma^\mu
q_2)|i(P_i)\rangle=\\
&&i\int\frac{d^4q_i}{(2\pi)^4}\mathrm{Tr}\Big[\frac{\bar{\eta}(q_{f_{P_f\perp}})\tilde{\Lambda}_{1}^{+}(q_{f_{P_{i\perp}}})
\Gamma^\mu\Lambda_{1}^{+}(q_{i_{P_{i\perp}}})\eta(q_{i_{P_{i\perp}}})\Lambda_{2}^{+}(q_{i_{P_{i\perp}}})}{(\alpha_{1f}P_{fP_{i}}+
q_{f_{P_{i}}}-\tilde{\omega}_1+i\epsilon)(\alpha_1M_i+q_{i_{P_i}}-\omega_1+i\epsilon)(\alpha_2M_i-q_{i_{P_i}}-\omega_2+i\epsilon)}+\dots\Big],\notag
\end{eqnarray}
where $\tilde{\Lambda}_{1}^{\pm}(q_{f_{P_{i\perp}}})\equiv
\frac{1}{2\tilde{\omega}_{1}}[\frac{\not\!{P}_i}{M_i}\tilde{\omega}_{1}\pm(\not\!{\tilde{p}_{1}}+m_{1f})]$
and
$\tilde{\omega}_{1(2)}\equiv\sqrt{m_{1(2)f}^2-\tilde{p}^2_{1(2)}}$
with $\tilde{p}_{1(2)}\equiv\alpha_{1(2)f}P_{fP_{i\perp}}\pm
q_{f_{P_{i\perp}}}$. The ``$\dots$" represent the terms involving
negative part ($\Lambda^-$).  Since the contributions of negative
energy parts are small, we can ignored them in equation
(\ref{eq2-ad2})~\cite{Chang}. Actually we compared $\varphi^{--}$
and $\frac{\Lambda^- \eta
\Lambda^+}{M_i-P_{fP_{i}}+\omega_{1f}+\omega_{1i}}$, which arise
from the terms with $\Lambda^-$('s), to $\varphi^{++}$ of $B_s$
meson numerically. The wave function of $B_s$ drops to $\sim 0$
when $|\vec{q}|\sim 2.1$~GeV, so the part with $|\vec{q}|>
2.1$~GeV makes rare contributions. It is found that for the main
part of the wave functions ($|\vec{q}|< 2.1$~GeV), $\varphi^{--}$
and $\frac{\Lambda^- \eta
\Lambda^+}{M_i-P_{fP_{i}}+\omega_{1f}+\omega_{1i}}$ are $\sim
0-2\%$ of $\varphi^{++}$ as $|\vec{q}|$ increases.

After integrating equation (\ref{eq2-ad2}) over $q_{i_{P_i}}$, one
obtains
\begin{equation}\label{eq2-7}
\langle f(P_f) |(\bar{q}_1\Gamma^\mu
q_2)|i(P_i)\rangle=\int\frac{d^3q_{i_{P_{i\perp}}}}{(2\pi)^3}\mathrm{Tr}\Big[\bar{\varphi}_{f}^{++}(q_{f_{P_{f\perp}}})
\frac{\not\!P_f}{M_f}L_{r}\Gamma^\mu\varphi_{i}^{++}(q_{i_{P_{i\perp}}})\frac{\not\!P_f}{M_f}\Big],
\end{equation}
where $\bar{\varphi}^{++}=\gamma_0\varphi^{++}\gamma_0$ and
$L_{r}=\frac{M_f-\omega_{1f}-\omega_{2f}}{P_{f_{P_i}}-\tilde{\omega}_1-\tilde{\omega}_2}\tilde{\Lambda}_{1}^{+}(q_{f_{P_{i\perp}}})$;
$q_{f_{P_{f\perp}}}=q_{f_{P_{i\perp}}}-\frac{q_{fP_{i\perp}}\cdot
P_{fP_{i\perp}}}{M_f^2} P_f+ s_r
(\frac{1}{M_i}P_i-\frac{P_{fP_{i}}}{M^2_f}P_f)$, with
$s_r=\alpha_{2f} P_{fP_{i}}-\omega_2$. In calculation, the
relation
$1=\frac{\not\!P_f}{M_f}\frac{\not\!P_f}{M_f}=(\Lambda_{1}^{+}(q_{f_{P_{f\perp}}})+\Lambda_1^-(q_{f_{P_{f\perp}}}))\frac{\not\!P_f}{M_f}$
has been used and again the negative part is ignored.  The
instantaneous transition matrix element used here, $i.e.$ equation
(\ref{eq2-7}), is different from the one in \cite{Wang2}. In
\cite{Wang2}, the instantaneous approximation is done in the
initial particle's rest frame for both the initial particle and
the final particle in the transition matrix element, whereas in
this method, the instantaneous approximation is done in the
relevant particle's own rest frame. It is found that, for the
transitions of $B_c$ decaying to charmed particles, the two
methods are consistent with each other; whereas for the
transitions of $B_s$ decaying to charmed particles, the former
generally gives larger form factors.

Recent years another method, in which the B-S equation also plays
an important role as in our method, was extensively
studied~\cite{DS1} and applied to describe meson
observables~\cite{DS2,DS3}. In the method (referred as the DSE
method for convenience), the rainbow-ladder truncation, which is a
symmetry-preserving truncation that grantee the axial-vector
vertices satisfying the Ward-Takahashi identity, is applied to the
kernel of the Dyson-Schwinger equation for a quark, i.e. the gap
equation, and the B-S equation's kernel. By solving the gap
equation and the B-S equation for considered channels, interested
observables could be estimated. The DSE method has impressed us
due to the success on describing the pion as both a Goldstone
mode, associated with dynamical chiral symmetry breaking (DCSB),
and a bound state composed of constituent $u$- and $d$-quarks. The
DSE method is mainly different from our method in two points.
First, the quark propagators are different; and second the B-S
kernel are different. The quark propagator in the DSE method, as
the solution of the gap equation, is the dressed quark propagator
which is characterized by a momentum-dependent mass function. In
our method, we use free quark propagators with constituent quark
masses. Of course, propagators appear in the B-S equation or
considered amplitudes should be the dressed one, however, studies
on the gap equation shown that the propagators of $u$, $d$, $s$
quarks receive strong momentum-dependent corrections at infrared
momenta while the mass functions in heavy quark $b$ and $c$
propagators can be approximated as a constant. Thus significant
differences appear in the light quark propagators. The propagators
in the DSE could also exhibit confinement characters of QCD. The
B-S kernel in the DSE method and in our method both have the one
gluon exchange interaction which is described by the products of
the strong running coupling constant and the free gluon
propagator. Except that we use an instantaneous kernel while the
DSE method does not, the differences of the kernels are, our
kernel also involves a confinement potential, while in the DSE
method the confinement is described by using quark propagators
with no Lehmann representation. The basic concepts of the DSE
method is attractive, however for applications involving a widely
ranges of observables, further assumptions and parameterization
are usually adopted~\cite{DS2}. For example, the B-S amplitude for
a heavy meson used in \cite{DS2} are obtained not by solving the
B-S equation, but by assuming a parameterized form and then
fitting the data to fix the parameters. Furthermore, the forms of
the B-S amplitudes of heavy mesons are too simple compared to
ours. Despite these differences in the propagators and the B-S
amplitudes, the form of a transition matrix element in the impulse
approximation in these works is the same as ours: the Mandelstam
formulation.

Due to the argument of Lorentz covariance, the transition matrix
element can be decomposed into several parts, where the form factors
show up. As usual, we denote form factors by the following
decompositions:
\begin{eqnarray}\langle
D_s^{-}
|V^\mu|B^0_s\rangle&\equiv& f_+(Q^2)P^\mu+f_-(Q^2)Q^\mu, \\
\langle D_s^{*-} |V^\mu|B^0_s\rangle&\equiv&
-i\frac{2}{M_i+M_f}f_V(Q^2)\varepsilon^{\mu\epsilon^* P_i
P_f},\\
\langle D_s^{*-} |A^\mu|B^0_s\rangle&\equiv&f_1(Q^2)\frac{\epsilon^*\cdot P_i}{M_i+M_f} P^\mu\notag\\&&
+f_2(Q^2)\frac{\epsilon^*\cdot P_i}{M_i+M_f} Q^\mu
+f_0(Q^2)(M_i+M_f)\epsilon^{*\mu},
\end{eqnarray}
where $P\equiv P_i+P_f$ and $Q\equiv P_i-P_f$. $f_\pm(Q^2)$,
$f_V(Q^2)$ and $f_i(Q^2)\  (i=0,1,2)$ are the form factors of weak
transition $B_{s}^{0}\rightarrow D_s^{(*)-}$.

\section{Non-leptonic two-body decay rate and its CP asymmetry for $\bar B_s^0\rightarrow D_s^{(*)+}X^-$}

In this section, we first treat the non-leptonic two-body decay in the
framework of the factorization approximation, and then introduce the pQCD method to estimate
the contributions from the non-factorizable effects and the annihilation diagrams. For
$\bar{B}^0_s\rightarrow D^{(*)+}_s+L^-$ decays induced by
$b\rightarrow c$ transition, where $L^-$ denotes a light meson, the
low energy effective weak Hamiltonian is given by
\begin{equation}\label{eq3-1}
\mathcal{H}_{\mathrm{eff}}=\frac{G_F}{\sqrt{2}}V_{cb}V^*_{uq}\Big\{C_1(\mu)Q_1+C_2(\mu)Q_2\Big\},
\end{equation}
where $Q_{1}=(\bar{c}_\alpha b_\alpha)_{V-A}(\bar{q}_\beta
u_\beta)_{V-A}$ and $Q_{2}=(\bar{c}_\alpha
b_\beta)_{V-A}(\bar{q}_\beta u_\alpha)_{V-A}$ with $q=d$ or $s$. And
for the double charmed $\bar{B}_s^0$ decays, the low energy
effective weak Hamiltonian for the $\Delta B=1$ transition
is~\cite{Buras},
\begin{eqnarray}\label{eq3-2}
\mathcal{H}_{\mathrm{eff}}(\Delta
B=1)=\frac{G_F}{\sqrt{2}}\sum\limits_{p=u,c}V_{pb}V^*_{pq}\left\{C_1(\mu)Q_1^p+C_2(\mu)Q_2^p
+\sum\limits_{i=3}^{10}C_i(\mu)Q_i\right\}+h.c.,
\end{eqnarray}
where $V_{pq}$ is the CKM matrix element with $(p=u,c)$ and
$(q=d,s)$. $C_i(\mu)$ are the Wilson coefficients. The local
four-quark operators $Q_i$ can be categorized into three groups: the
tree operators in $b\rightarrow p$ transition $Q_1^p,\ Q_2^p$, the
QCD penguin operators $Q_i\ (i=3,4,5,6)$, and the electroweak
penguin operators $Q_i\ (i=7,8,9,10)$. All these local four-quark
operators are written as
\begin{equation}\begin{aligned}\label{eq3-3}
Q_1^p=(\bar{q}_\alpha p_\alpha)_{V-A}(\bar{p}_\beta
b_\beta)_{V-A},\\
Q_2^p=(\bar{q}_\alpha p_\beta)_{V-A}(\bar{p}_\beta
b_\alpha)_{V-A},
\end{aligned}
\end{equation}

\begin{equation}\begin{aligned}\label{eq3-4}
Q_3=(\bar{q}_\alpha b_\alpha)_{V-A}\sum_{q_x}(\bar{q}_{x\beta}
q_{x\beta})_{V-A},\\
Q_4=(\bar{q}_\alpha b_\beta)_{V-A}\sum_{q_x}(\bar{q}_{x\beta}
q_{x\alpha})_{V-A},\\
Q_5=(\bar{q}_\alpha b_\alpha)_{V-A}\sum_{q_x}(\bar{q}_{x\beta}
q_{x\beta})_{V+A},\\
Q_6=(\bar{q}_\alpha b_\beta)_{V-A}\sum_{q_x}(\bar{q}_{x\beta}
q_{x\alpha})_{V+A},
\end{aligned}
\end{equation}

\begin{equation}
\begin{aligned}\label{eq3-5}
Q_7=\frac{3}{2}(\bar{q}_\alpha
b_\alpha)_{V-A}\sum_{q_x}e_{q_x}(\bar{q}_{x\beta}
q_{x\beta})_{V+A},\\
Q_8=\frac{3}{2}(\bar{q}_\alpha
b_\beta)_{V-A}\sum_{q_x}e_{q_x}(\bar{q}_{x\beta}
q_{x\alpha})_{V+A},\\
Q_9=\frac{3}{2}(\bar{q}_\alpha
b_\alpha)_{V-A}\sum_{q_x}e_{q_x}(\bar{q}_{x\beta}
q_{x\beta})_{V-A},\\
Q_{10}=\frac{3}{2}(\bar{q}_\alpha
b_\beta)_{V-A}\sum_{q_x}e_{q_x}(\bar{q}_{x\beta} q_{x\alpha})_{V-A},
\end{aligned}
\end{equation}
where $q_x$ ranges from $u,d,s$ to $c$. The subscripts
$\alpha,\beta$ are color indices. The operator
$(\bar{\psi}_{1\alpha} \psi_{2\beta})_{V\pm A}$
$\equiv\bar{\psi}_{1\alpha}\gamma^\mu(1\pm\gamma_5) \psi_{2\beta}$.
As usual, we define the combinations $a_i$ of Wilson coefficients
\begin{equation}a_{2i-1}\equiv C_{2i-1}+\frac{C_{2i}}{N_c},\ \  a_{2i}\equiv C_{2i}+\frac{C_{2i-1}}{N_c},\end{equation}
where $N_c$ is the number of quark colors and is taken as $N_c=3$.

Under the FA, the matrix element of $\bar{B}^0_s\rightarrow
D_s^{+(*)}X^-$ two body decays can be factorized as~\cite{Kim,Bauer}
\begin{equation}A=\langle D^{(*)+}_s
|(\bar{c}b)_{V-A}|\bar{B}^0_s\rangle\langle X^-
|(\bar{q}p)_{V-A}|0\rangle,\end{equation}
where $\langle X^-(P_X)
|(\bar{q}p)_{V-A}|0\rangle\equiv if_{0^{\pm}} P_{X}^\mu$ when the
meson $X$ denotes a scalar (pseudoscalar), and $\langle X^-(P_X)
|(\bar{q}p)_{V-A}|0\rangle\equiv if_{1^{\pm}}M_X\epsilon^{*\mu}$
when $X$ is a axial vector (vector). $f_{0^\pm}$ and $f_{1^\pm}$ are
decay constants of particle $X$. The decay amplitudes for the
$\bar{B}^0_s\rightarrow D^{(*)+}_s+L^-$ decays can be expressed as
\begin{equation}\label{eq3-6}
\mathcal{M}=\frac{G_F}{\sqrt{2}} V_{cb}V^*_{uq} a_1 A.
\end{equation}
The decay amplitude of double charmed $\bar{B}^0_s$ decay can be
written as~\cite{Kim}
\begin{equation}\label{eq3-6}
\mathcal{M}=\frac{G_F}{\sqrt{2}}\Big\{\lambda_c
a_1+\sum\limits_{p=u,c}\lambda_p\big[a^p_4+a^p_{10}+\xi(a^p_6+a^p_8)\big]\Big\}A,
\end{equation}
where $\lambda_p\equiv V_{pb}V^*_{pq}$ and $a_i^p\equiv a_i+I_i^p$
with $I_i^p$ given as follows:
\begin{equation}I_4^p=I_6^p=\frac{\alpha_s}{9\pi}\big\{C_1[\frac{10}{9}-G(m_p,k^2)]\big\},\end{equation}
\begin{equation}I_8^p=I_{10}^p=\frac{\alpha_e}{9\pi}\frac{1}{N_c}\big\{(C_1+C_2N_c)[\frac{10}{9}-G(m_p,k^2)]\big\}.\end{equation}
The penguin loop integral function $G(m_p,k^2)$ is given by
\begin{equation}\label{eq3-7}
G(m_p,k^2)=-4\int_0^1x(1-x)\mathrm{ln}\frac{m_{p}^2-k^2x(1-x)}{m_b^2}dx,
\end{equation}
where the penguin momentum transfer
$k^2=\frac{m_b^2}{2}(1+(m_{\bar{q}_x}^2-m_{q}^2)(1-\frac{m_{\bar{q}_x}^2}{m_b^2})/M_X^2
+(m_{q}^2+2m_{\bar{q}_x}^2-M_X^2)/m_b^2)$~\cite{Du1}. The $\xi$ in
equation (\ref{eq3-6}) arises from the contribution of the
right-handed currents and depends on the $J^P$ quantum numbers of
the final state particles. The collected expressions of $\xi$ are
shown as follows:
\begin{equation}\label{eq3-8}
\xi=\left\{
\begin{array}{cc}+\frac{2M_X^2}{(m_b-m_c)(m_c+m_q)}~, ~~~~&D_sX(0^-)\\
-\frac{2M_X^2}{(m_b-m_c)(m_q-m_c)}~, ~~~~&D_sX(0^+)\\
-\frac{2M_X^2}{(m_b+m_c)(m_c+m_q)}~, ~~~~&D^*_sX(0^-)\\
+\frac{2M_X^2}{(m_b+m_c)(m_q-m_c)}~, ~~~~&D^*_sX(0^+)\\
0~, ~~~~~~&D^{(*)}_sX(1^{\pm})
\end{array}\right.
\end{equation}
where $X$ denotes a $D_q\ (q=s,d)$ meson with its $J^P$ shown in
the bracket just following it. The current quark masses
encountered in $G(m_p,k^2)$ and $\xi$ are taken from \cite{PDG}
and then evolved to the scale $\mu\sim m_b$ by the renormalization
group equation of the running quark masses~\cite{Buras}:
\begin{equation}
m(\mu)=m(\mu_0)\left\{\frac{\alpha_s(\mu)}{\alpha_s(\mu_0)}\right\}^{\frac{\gamma_{m0}}{2\beta_0}}
\left\{1+(\frac{\gamma_{m1}}{2\beta_0}-\frac{\beta_1\gamma_{m0}}{2\beta_0^2})\frac{\alpha_s(\mu)-\alpha_s(\mu_0)}{4\pi}\right\},
\end{equation}
where \begin{eqnarray}\beta_0=\frac{11N_c-2f}{3},&&
\beta_1=\frac{34}{3}N_c^2-\frac{10}{3}N_cf-2C_Ff,\notag\\
\gamma_{m0}=6C_F,&&
\gamma_{m1}=C_F(3C_F+\frac{97}{3}N_c-\frac{10}{3}f),\end{eqnarray}
and $C_F=\frac{N_c^2-1}{2N_c}$. The number of quark flavors is denoted as $f$,
which is taken as $f=5$ in the present paper.

As indicated before, under the FA non-factorizable effects and the
annihilation contributions are both neglected.  But for double
charmed $B_s$ decays they may contribute conspicuously, especially
for CPAs, since it has been indicated that the annihilation
diagrams usually make domain contribution on the strong phases
according to the pQCD analysis. Thus in this work, we estimate the
non-factorizable and annihilation contributions in the pQCD
approach to make more reliable predictions on non-leptonic decays.
Concretely, we will estimate the contributions from
non-factorizble and annihilation (if exits) for
$\bar{B}^0_s\rightarrow D^{(*)+}_s+X^-$ decays, where only
pseudoscalar or vector present in the final state. Here we will
not illustrate the pQCD approach in detail, instead we refer the
readers to the original paper~\cite{PQCD} and \cite{Li2} which
provides the specific pQCD studies on $\bar{B}^0_s\rightarrow
D^{(*)+}_s+X^-$ decays for details of this method. Thanks to the
efforts did by the authors of \cite{Li2}, from which we borrow the
expressions of decay amplitudes of non-factorizable and
annihilation diagrams. But we use our form factors in estimating
the factorizable color-favored diagrams' contributions. For
example, the decay amplitude of the $\bar{B}^0_s\rightarrow
D^{(*)+}_sD^-$ decay could be expressed as
\begin{eqnarray}
\mathcal{M}&=&\frac{G_F}{\sqrt{2}}\Big\{V_{cb}V^*_{cd}\left[F_{e}^{LL}(a_1)+F_{en}^{LL}(C_1)+S_F\right]+V_{ub}V^*_{ud}S_F)\Big\},\\
S_F&=&F_{e}^{LL}(a_4+a_{10})+F_{en}^{LL}(C_3+C_9)+F_{e}^{SP}(a_6+a_8)+F_{en}^{LR}(C_5+C_7)\notag\\&&
+F_{a}^{LL}(a_4-a_{10}/2)+F_{an}^{LL}(C_3-C_9/2)+F_{a}^{SP}(a_6-a_8/2)+F_{an}^{LR}(C_5-C_7/2),\notag
\end{eqnarray}
where each $F_{e(n),a(n)}^{LL(LR,SP)}(\dots)$ corresponds to a
certain diagram's contribution. The subscript ``e" represents
factorizable emission (color-favored) diagrams; ``en" represents
non-factorizable emission diagrams; ``a" and ``an" represents
factorizable and non-factorizable annihilation diagrams
respectively. The superscript ``LL", ``LR" and ``SP" correspond to
the contributions from the (V-A)(V-A) operators, the (V-A)(V+A)
operators and (S-P)(S+P) operators respectively. In this work, we
calculate the factorizable emission contributions, which can be
expressed in terms of form factors and decay constants, with our
estimated form factors; and calculate the other contributions with
the pQCD approach. The exact expressions of $F_{e(n)}, F_{a(n)}$
can be found in \cite{Li2}.

The decay width of a two-body decay is
\begin{equation}\label{eq3-9}
\Gamma=\frac{|\vec{p}|}{8\pi
M_{\bar{B}_s^0}^2}\sum_\mathrm{pol}|\mathcal{M}|^2,
\end{equation} where
$\vec{p}$ is the 3-momentum of one of the final state particles in
the rest frame of $\bar{B}_s^0$. Another important physical
observable is CP asymmetry. Generally the amplitude for double
charmed $\bar{B}_s^0$ decays considered here can be written as
$\mathcal{M}=V_{cb}V^*_{cq}T_1+V_{ub}V^*_{uq}T_2.$ The direct CP
asymmetry arise from the interference between the two parts of the
amplitude and is defined as
\begin{equation}\begin{aligned}\label{eq3-12}
\mathcal{A}^{\mathrm{dir}}_{cp}&\equiv\frac{\Gamma(B_s^0\rightarrow
f)-\Gamma(\bar{B}_s^0\rightarrow \bar{f})}{\Gamma(B_s^0\rightarrow
f)+\Gamma(\bar{B}_s^0\rightarrow \bar{f})}\\
&=\frac{\epsilon_i2\sin\delta\sin\gamma}
{|G_1/G_2|+|G_2/G_1|+\epsilon_i2\cos\delta\cos\gamma}\\
&= D_1\frac{\sin\gamma}{1+D_2\cos\gamma},
\end{aligned}\end{equation}
where the weak phase
$\gamma\equiv\arg(-\frac{V^*_{ub}V_{ud}}{V^*_{cb}V_{cd}})\simeq\arg(\frac{V^*_{ub}V_{us}}{V^*_{cb}V_{cs}})$,
the strong phase $\delta=\arg(T_1)-\arg(T_2)$,
$G_1=V_{cb}V^*_{cq}T_1$, $G_2=V_{ub}V^*_{uq}T_2$ and
\begin{equation}\label{eq3-13}
D_1\equiv\frac{\epsilon_i2\sin\delta}{|G_1/G_2|+|G_2/G_1|},~~~~
D_2\equiv\frac{\epsilon_i2\cos\delta}{|G_1/G_2|+|G_2/G_1|},
\end{equation}
with $\epsilon_1=+1$ for $q=s$, and $\epsilon_2=-1$ for $q=d$,
respectively.

Besides the branching ratios and CP asymmetries, the polarization
fraction of $B_s\rightarrow VV(A)$ decays is another important
observable. To illustrate the polarization fraction, one can write
the decay amplitude as~\cite{CGW}
\begin{equation}\mathcal{M}=f_{1^\pm}M_{V(A)2}[a~ P_i^\mu P_i^\nu+ b~
g^{\mu\nu}+ i c ~\varepsilon^{\nu\mu P_i
P_{V1}}]\epsilon^*_{V1\mu}(\lambda_1)\epsilon^*_{V(A)2\nu}(\lambda_2),\end{equation}
where $P_i~(M_i)$, $P_{V1}~(M_{V1})$ and $P_{V(A)2}~(M_{V(A)2})$ are
the momenta (masses) of the initial particle, the particle picking
up the spectator quark in the final state and the other meson in the
final state, respectively. Coefficients $a,~b$ and $c$ are defined
as $a=2\tilde{C}f_1/(M_i+M_{V1})$, $b=f_0\tilde{C}(M_i+M_{V1})$ and
$c=2\tilde{C}f_V/(M_i+M_{V1})$, where $\tilde{C}$ denotes the term
involving coupling constant, relevant Wilson coefficients and CKM
matrix elements in front of the hadron matrix element $A$.
$\lambda's$ are the helicities of the final particles. Then the
decay amplitude of various helicities can be given as
\begin{equation}\mathcal{M}_L=f_{1^\pm}\left[\frac{a~
M_i^2\vec{P}_{V1}^2+b~(\vec{P}_{V1}^2+P^0_{V1}P^0_{V(A)2})}{M_{V1}}\right],\end{equation}
\begin{equation}\mathcal{M}_{\parallel}=\sqrt{2}~b~f_{1^\pm}M_{V(A)2},~~~~\mathcal{M}_{\perp}=\sqrt{2}~c~f_{1^\pm}M_{V(A)2}M_i|\vec{P}_{V1}|,\end{equation}
where $\mathcal{M}_L$, $\mathcal{M}_\parallel$ and
$\mathcal{M}_\perp$ denote longitudinal, transverse parallel and
transverse perpendicular part of the amplitude, respectively. The expressions of $\mathcal{M}$'s apply under the FA.
For pQCD calculations certain terms corresponding to non-factorizable and annihilation diagrams
should be added to each $\mathcal{M}$. The
momentum $\vec{P}_{V1}$ and energy $P^0_{V1}(P^0_{V(A)2})$ are taken
in the rest frame of the initial particle, $i.e.$ $B_s^0$. The
polarization fraction is defined as
$R_i=\frac{|\mathcal{M}_i|^2}{|\mathcal{M}_L|^2+|\mathcal{M}_{\parallel}|^2+|\mathcal{M}_{\perp}|^2}$,
where $i=L$, $\parallel$ and $\perp$.

\section{Numerical results and discussions}

\subsection{Form Factors of $B^0_s\rightarrow D_s^{(*)-}$ Transition and Semi-leptonic decays}

\begin{figure}[t]
\centering
\includegraphics[width = 0.8\textwidth]{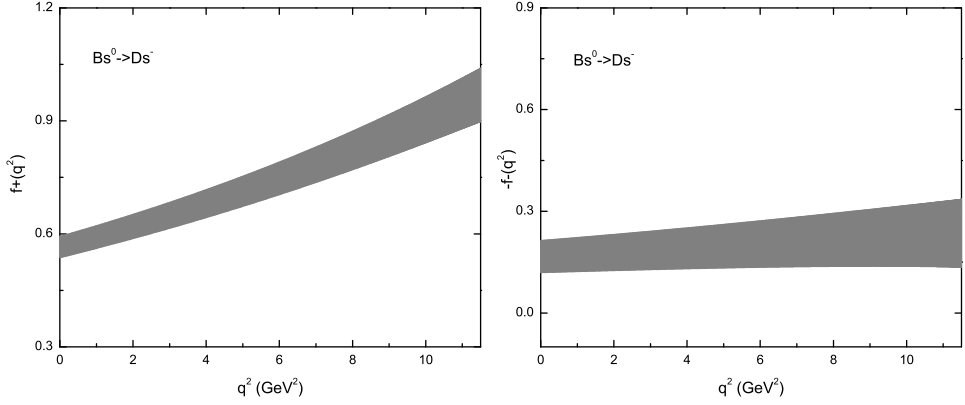}
\caption{Form factors of $B^0_s\rightarrow D_s^{-}$ weak
transition.} \label{fig4-1}
\end{figure}

\begin{figure}[t]
\centering
\includegraphics[width = 0.8\textwidth]{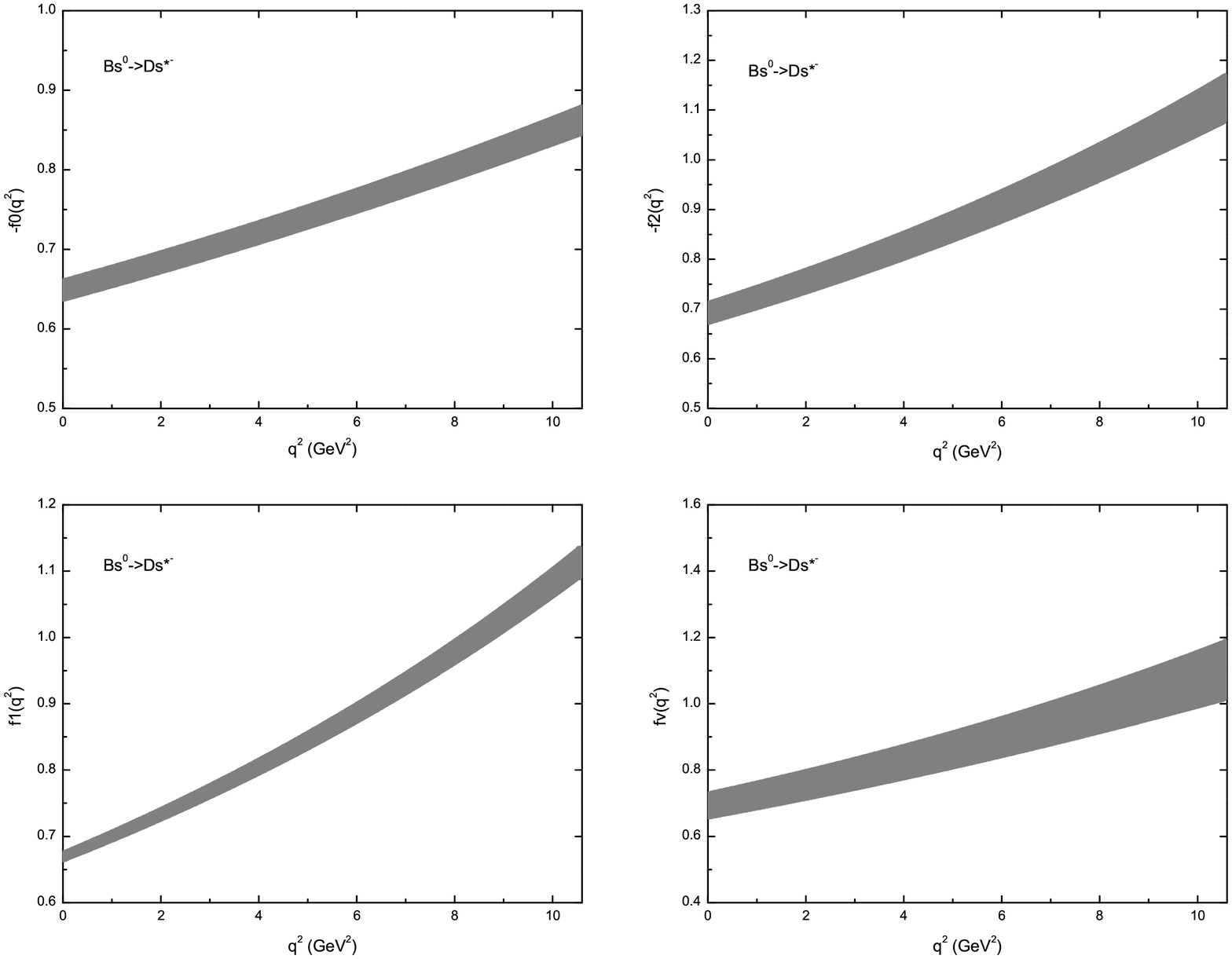}
\caption{Form factors of $B^0_s\rightarrow D_s^{*-}$ weak
transitions.} \label{fig4-2}
\end{figure}

By solving the Salpeter equation~(\ref{eq2-n1}), we obtain the
wave functions of  $B_s^0$, $D_s^{\pm}$, $D_s^{*\pm}$ mesons. Then
we calculate the form factors of $B^0_s\rightarrow D_s^{(*)-}$
transition in the whole physical region numerically with equation
(\ref{eq2-7}). In calculation, the particles' masses
$M_{B_s^0}=5366.3$~MeV, $M_{D_s^{\pm}}=1968.47$~MeV and
$M_{D_s^{*\pm}}=2112.3$~MeV~\cite{PDG} are used. The results are
drawn in Fig.~\ref{fig4-1} and Fig.~\ref{fig4-2} for $D_s^-$ and
$D_s^{*-}$, respectively. The parameter-dependent uncertainty can
be estimated by varying the input parameters of our model in a
reasonable range. In this work we vary the parameters $m_b$,
$m_c$, $m_s$, $\lambda$ and $\Lambda_{\mathrm{QCD}}$ by $\pm5\%$
to give the errors. In this section only, we denote the momentum
transfer $q\equiv P_i-P_f$, instead of $Q$. It should be noticed
that our theoretical estimations suffer other uncertainties
arising from the instantaneous approximation, since the $s$-quark
is not a heavy quark. It is well known that describing the
inter-quark interactions by a QCD-inspired potential (which
relates to the instantaneous kernel) works well for a meson
consisting of a heavy quark and a heavy anti-quark (i.e.
$b(\bar{b})$, $c(\bar{c})$), but is questionable in describing
light mesons. For the present case, as $s$-quark is not heavy
enough, assuming the (anti-)quarks interact instantaneously may
cause (maybe sizeable) uncertainties, which also means that
retardation effects may give contributions (maybe sizeable). Till
now the problem encountered here is not totally solved. But part
of the retardation effects have been studied in \cite{Ret}. In
that work, the authors assume the confinement kernel as
$V_s\sim1/(-q_0^{2}+\vec{q}^2)^2$ and approximate it to be
$\frac{1}{\vec{q}^4}(1+2q_0^2/\vec{q}^2)$ by expanding
$\frac{q_0^2}{\vec{q}^2}$. Then the $q_0$ is replaced by its
``on-shell" value, which are obtained by assuming that quarks are
on their mass shells. The ``on-shell" approximation imply that the
considered meson should be a weak binding system. Finally, the
total effect is adding a term
$\frac{2\lambda}{\pi^2(\vec{q}^2+a^2)^3}(\sqrt{(\vec{q}-\vec{k})^2+m^2}-\sqrt{\vec{k}^2+m^2})^2$
to $V_s$. $m$ and $\vec{k}$ are the constituent mass and momentum
of the (anti-)quark. For heavy-light system, it is better to take
$m$ and $\vec{k}$ to be the heavy quark's mass and momentum. By
using such a interaction kernel, some of the retardation effects
could be incorporated in calculations. Of course, this method
didn't solve the problem totally, because only some of the
retardation effects are incorporated and we don't know how much
they are. Thus we won't take this interaction kernel as a
corrected version of our potential in equation (\ref{eq2-ad7}).
But in order to obtain a qualitative feeling about the
uncertainties arising from instantaneous approximation, we use the
interaction kernel presented in \cite{Ret}, which incorporated
some retardation effects, to estimate the mass spectra and form
factors and compare them to our results without retardation
effects. It is found that the relative variations between the two
sets of results are: $\frac{\Delta M_{B_s}}{M_{B_s}} \sim0.3\%$,
$\frac{\Delta M_{D_s{(*)}}}{M_{D_s{(*)}}} \sim5\%$ and
$\frac{\Delta f_i(q^2=0)}{f_i(q^2=0)} \sim 4-7\%$. Due to the
reason indicated before, we emphasize that the actual errors
(caused by describing the inter-quark interaction with a
potential) may be larger.

In
Table~\ref{tab4-1}, we compare our form
factors at $q^2=0$ with those from other approaches. This can be seen from the table: for
$B_s^0\rightarrow D_s^-$ transition, our results are roughly
consistent with the results of BSW model and QCDSR, but larger
than those of LCSR, therefore, it is expected that the LCSR method
may give smaller decay rates for $B_s^0\rightarrow D_s^-+L^+$
non-leptonic decays in which the momentum transfer is near $q^2=0$
GeV$^2$. For $B_s^0\rightarrow D_s^{*-}$ transition, our results
are a little larger than the results from other methods.

\begin{table} \caption{Form factors of $B^0_s\rightarrow D_s^{-}$ and $B^0_s\rightarrow D_s^{*-}$
transitions at $q^2=0$~GeV$^2$.}
\begin{center}\begin{tabular}{|c|c|c|c|c|c|} \hline\hline &This work& BSW~\cite{Kramer}&
QCDSR~\cite{Blasi}&LCSR~\cite{Li1}&CLFQM~\cite{GangLi}\\\hline
$f_{+}(0)$ &$0.57^{+0.02}_{-0.03}$&0.61&$0.7\pm0.1$&0.43&\\
\hline$f_{-}(0)$&$-0.17^{+0.05}_{-0.04}$&&&$-0.17$&\\\hline
$f_{V}(0)$ &$0.70^{+0.03}_{-0.04}$&0.64&$0.63\pm0.05$&&$0.74\pm0.05$\\
\hline$f_{0}(0)$&$-0.65^{+0.01}_{-0.01}$&$-0.56$&$-0.62\pm0.01$&&$-0.61\pm0.03$\\\hline
$f_{1}(0)$ &$0.67^{+0.01}_{-0.01}$&0.59&$0.75\pm0.07$&&$0.59\pm0.04$\\
\hline$f_{2}(0)$&$-0.70^{+0.03}_{-0.01}$&&&&\\\hline
\end{tabular}\end{center}\label{tab4-1}
\end{table}

For semi-leptonic $B_s^0\rightarrow D_s^{(*)-}l^+\nu_l$ decays
induced by $\bar{b}\rightarrow \bar{c}\nu_l\bar{l}$ transition, the
effective Hamiltonian can be written as~\cite{Li1}
\begin{equation}\label{eq4-ad1}
\mathcal{H}_{\mathrm{eff}}=\frac{G_F}{\sqrt{2}}V^*_{cb}\bar{b}\gamma_\mu(1-\gamma_5)c\bar{\nu}_l\gamma^\mu(1-\gamma_5)l.
\end{equation}
The amplitude of $B_s^0\rightarrow D_s^{(*)-}l^+\nu_l$ decays
could be obtained by sandwiching equation (\ref{eq4-ad1}) between
the initial and final states, which reads
\begin{equation}\label{eq4-ad2}
\mathcal{M}=\frac{G_F}{\sqrt{2}}V^*_{cb}\bar{\nu}_l\gamma^\mu(1-\gamma_5)l\langle
D^{(*)-}_s |\bar{b}\gamma_\mu(1-\gamma_5)c|B_s^0\rangle.
\end{equation}
The width of a semi-leptonic decay is
$\Gamma=\frac{1}{8M_{B_s}(2\pi)^3}\int\sum_\mathrm{pol}|\mathcal{M}|^2dE_ldE_f$
where $E_l$ and $E_f$ are the energies of the lepton and the meson
in the final state respectively. With the form factors calculated,
we estimate the branching ratios of semi-leptonic
$B_s^0\rightarrow D_s^{(*)-}l^+\nu_l$ decays. The results are
listed in Table~\ref{tab4-ad1} together with those from other
approaches. The orders of magnitude for those branching ratios are
consistent with each other. The $B_s^0\rightarrow
D_s^{(*)-}e^+\nu_e$ decay rates have been studied in the same
model in ~\cite{Zhang} as used here. But the parameters and the
formulation of the transition matrix elements used in that
reference are different from those in this work, thus their
results are larger. The parameters used in calculation such as
$V_{cb}$, $\tau_{B_s^0}$ are shown in the next subsection.

\begin{table} \scriptsize\caption{Branching ratios for semi-leptonic $B_s^0\rightarrow D_s^{(*)-}l^+\nu_l$
decays compared with those from other approaches.}
\begin{center}\begin{tabular}{|c|c|c|c|c|c|}
\hline\hline Processes&This work&CQM~\cite{CQM}&QCDSR~\cite{Azizi}&LCSR~\cite{Li1}&CLFQM~\cite{GangLi} \\
\hline $B_s^0\rightarrow D_s^{-}l^+\nu_l~(l=e,\mu)$&$(1.4-1.7)~\%$&$(2.73-3.00)\%$&$(2.8-3.8)\%$&$1.0^{+0.4}_{-0.3}~\%$&\\
\hline $B_s^0\rightarrow
D_s^{-}\tau^+\nu_{\tau}$&$(4.7-5.5)\times10^{-3}$&&&$3.3^{+1.4}_{-1.1}\times10^{-3}$&\\\hline
\hline $B_s^0\rightarrow
D_s^{*-}l^+\nu_l~(l=e,\mu)$&$(5.1-5.8)~\%$&$(7.49-7.66)\%$&$(1.89-6.61)\%$&&$5.2^{+0.6}_{-0.6}~\%$\\
\hline $B_s^0\rightarrow
D_s^{*-}\tau^+\nu_{\tau}$&$(1.2-1.3)~\%$&&&&$1.3^{+0.2}_{-0.1}~\%$\\\hline
\end{tabular}\end{center}\label{tab4-ad1}
\end{table}

\subsection{Non-leptonic $B_s$ decay}

Now we can use the form factors to estimate the decay rates of
$B_s^0$. The CKM matrix elements used in our calculation are~\cite{PDG}
\begin{equation}\begin{array}{ccc}
|V_{ud}|=0.97425,&|V_{us}|=0.2252,&|V_{ub}|=3.89\times10^{-3},\notag\\
|V_{cd}|=0.230,&|V_{cs}|=0.9735,&|V_{cb}|=0.0406.
\end{array}
\end{equation}
The lifetime $\tau_{B_s^0}=1.472\times10^{-12}s$~\cite{PDG} is
taken in calculation. The Wilson coefficients are quoted from
\cite{Sun}, and for $\mu\sim m_b$, they are:
$$\begin{array}{ccccc}
C_1=1.0849,&C_2=-0.1902,&C_3=0.0148,&C_4=-0.0362,&C_5=0.0088,\\
C_6=-0.0422,&\frac{C_7}{\alpha_{e}}=-0.0007,&\frac{C_8}{\alpha_{e}}=0.0565,&\frac{C_9}{\alpha_{e}}=-1.3039,&\frac{C_{10}}{\alpha_{e}}=0.2700,\\
\end{array}$$
where $\alpha_e(M_W)=1/128$ is the electromagnetic coupling
constant. The strong coupling constant is taken as
$\alpha_s(m_b)=0.216$. The decay constants used in this paper are
shown in Table~\ref{tab4-2}. Other inputs in pQCD analysis such as
the wave functions and the Jet function appearing in the
non-factorizable amplitudes and annihilation amplitudes are taken
as the same as in \cite{Li2}.

\begin{table} \caption{Decay constants used in our calculation in unit of MeV.}
\begin{center}\begin{tabular}{|c|c|c|c|c|c|c|}
\hline\hline$f_{\pi}$&$f_{k}$&$f_{D}$&$f_{D_s}$&$f_{\rho}$&$f_{k^*}$
&$f_{a_1}$\\\hline 130 \cite{PDG}&156 \cite{PDG}&$207^{+9}_{-9}$
\cite{PDG}&$258^{+6}_{-6}$ \cite{PDG}&$205^{+9}_{-9}$
\cite{Ball}&$217^{+5}_{-5}$ \cite{Ball}&229
\cite{Rosner}\\\hline\hline
$f_{D^*}$&$f_{D_s^*}$&$f_{D^*_0}$&$f_{D_{s0}^*}$&$f_{D_{s1}(2460)}$&$f_{D_{s1}(2536)}$
&$f_{B_s}$\\\hline $245^{+20}_{-20}$ \cite{Bec}&$272^{+16}_{-16}$\cite{Bec}
&$137^{+4}_{-5}~\cite{Fu}$&$109^{+4}_{-5}~\cite{Fu}$&$227^{+22}_{-19}~\cite{Fu}$&$77.3^{+12.4}_{-9.8}~\cite{Fu}$&0.23~\cite{Li2}\\\hline
\end{tabular}\end{center}\label{tab4-2}
\end{table}

With these input parameters, we calculate the branching ratios of
non-leptonic $\bar{B}_s^0$ to charmed particle decays. The results
are listed in Table~\ref{tab4-3}, together with those from other methods, as
well as with available experimental data. The branching ratios of
double charmed decays shown in the table are the CP averaged values:
$\frac{1}{2}(\mathcal{B}(B_s^0\rightarrow
f)+\mathcal{B}(\bar{B}_s^0\rightarrow \bar{f}))$, where
$\mathcal{B}$ denotes branching ratio. The momentum transfer of
$\bar{B}^0_s\rightarrow D_s^{(*)+}L^-$ decays, where $L^-$ is a
light meson, is $q^2\sim 0-1.5~\mathrm{GeV}^2$, and the momentum
transfer of the double charmed decays is
$q^2\sim3.5-6.4~\mathrm{GeV}^2$.

\begin{table} \tiny\caption{The (averaged) branching ratios of $\bar{B}^0_s\rightarrow D_s^{(*)+}X^-$
in the unit of $\%$. Our results with pQCD represent those with
non-factorziable and annihilation contributions estimated in the
pQCD approach. The parameters in \cite{Bijnens,Blasi,Deandrea} are
replaced by the ones used in our paper: $|V_{cb}|=0.04$,
$\tau_{B_s^0}=1.472\times 10^{-12}$s, $a_1=1.02$. The results of
$\bar{B}^0_s\rightarrow D^{+}_sX^-$ in the 7th column are quoted
from \cite{Li1} and the results of $\bar{B}^0_s\rightarrow
D^{*+}_sX^-$ in the 7th column are quoted from \cite{GangLi}.
Experimental data are quoted from \cite{PDG,Belle}.}
\begin{center}\begin{tabular}{|c|c|c|c|c|c|c|c|c|}
\hline\hline channels&Ours (FA) &Ours (pQCD) &
BSW~\cite{Bijnens}&QCDSR~\cite{Blasi}&HQET~\cite{Deandrea}&\cite{GangLi}~\cite{Li1}&pQCD~\cite{Li2}&
Experimental Data\\\hline
$D_s^+\pi^-$&$0.27^{+0.02}_{-0.03}$&$0.27^{+0.02}_{-0.03}$&0.37&0.41&0.31&$0.17^{+0.07}_{-0.06}$&$0.196^{+0.106}_{-0.075}$&$0.32\pm0.05$\\\hline
$D_s^+\rho^-$&$0.64^{+0.12}_{-0.11}$&$0.60^{+0.11}_{-0.11}$&0.94&1.1&0.82&$0.42^{+0.17}_{-0.14}$&$0.47^{+0.249}_{-0.177}$&$0.85^{+0.13}_{-0.12}\pm0.11\pm0.13$\\\hline
$D_s^+k^-$&$0.021^{+0.002}_{-0.002}$&$0.021^{+0.001}_{-0.002}$&0.028&0.033&0.023&$0.013^{+0.005}_{-0.004}$&$0.0170^{+0.0087}_{-0.0066}$&\\\hline
$D_s^+k^{*-}$&$0.038^{+0.005}_{-0.005}$&$0.038^{+0.005}_{-0.005}$&0.049&0.049&0.040&$0.024^{+0.010}_{-0.008}$&$0.0281^{+0.0147}_{-0.0109}$&\\\hline
$D_s^+a_1^-$&$0.75^{+0.06}_{-0.08}$&&0.88&0.9&&&&\\\hline\hline
$D_s^{*+}\pi^-$&$0.31^{+0.03}_{-0.02}$&$0.30^{+0.03}_{-0.03}$&0.28&0.16&0.31&$0.35^{+0.04}_{-0.04}$&$0.189^{+0.103}_{-0.072}$&$0.24^{+0.05}_{-0.04}\pm0.03\pm0.04$\\\hline
$D_s^{*+}\rho^-$&$0.90^{+0.15}_{-0.15}$&$0.91^{+0.16}_{-0.15}$&0.88&1.1&0.97&$1.18^{+0.33}_{-0.31}$&$0.523^{+0.283}_{-0.195}$&$1.19^{+0.22}_{-0.20}\pm0.17\pm0.18$\\\hline
$D_s^{*+}k^-$&$0.024^{+0.002}_{-0.002}$&$0.024^{+0.001}_{-0.003}$&0.020&0.016&0.023&$0.028^{+0.003}_{-0.003}$&$0.0164^{+0.0084}_{-0.0064}$&\\\hline
$D_s^{*+}k^{*-}$&$0.056^{+0.006}_{-0.007}$&$0.058^{+0.007}_{-0.007}$&0.048&0.049&0.053&$0.055^{+0.006}_{-0.006}$&$0.0322^{+0.0183}_{-0.0124}$&\\\hline
$D_s^{*+}a_1^-$&$1.3^{+0.1}_{-0.1}$&&1.1&&&&&\\\hline\hline
$D_s^+D^-$&$0.031^{+0.006}_{-0.005}$&$0.036_{-0.006}^{+0.007}$&0.027&0.041&0.049&$0.011^{+0.004}_{-0.004}$&$0.022^{+0.014}_{-0.010}$&\\\hline
$D_s^+D^{*-}$&$0.034^{+0.009}_{-0.008}$&$0.041^{+0.011}_{-0.011}$&0.031&0.033&0.034&$0.014^{+0.006}_{-0.005}$&$0.021^{+0.013}_{-0.009}$&\\\hline
$D_s^+D_{0}^{*-}$&$0.012^{+0.002}_{-0.002}$&&&&&$0.002^{+0.001}_{-0.001}$&&\\\hline
$D_s^{*+}D^-$&$0.032^{+0.006}_{-0.006}$&$0.038^{+0.006}_{-0.008}$&0.013&0.016&0.036&$0.037^{+0.004}_{-0.004}$&$0.027^{+0.017}_{-0.011}$&\\\hline
$D_s^{*+}D^{*-}$&$0.11^{+0.02}_{-0.03}$&$0.13^{+0.03}_{-0.03}$&0.074&0.065&0.11&$0.086^{+0.010}_{-0.009}$&$0.039^{+0.026}_{-0.019}$&\\\hline
$D_s^{*+}D_{0}^{*-}$&$0.010^{+0.002}_{-0.001}$&&&&&&&\\\hline\hline
$D_s^+D_s^-$&$0.83^{+0.10}_{-0.10}$&$1.3^{+0.1}_{-0.3}$&0.51&0.82&1.2&$0.35^{+0.14}_{-0.12}$&$0.55^{+0.36}_{-0.24}$&$1.04\pm0.35$\\\hline
$D_s^+D_s^{*-}$&$0.70^{+0.16}_{-0.15}$&$1.0^{+0.2}_{-0.2}$&0.57&0.65&0.77&$0.33^{+0.13}_{-0.11}$&$0.48^{+0.31}_{-0.21}$&\\\cline{1-8}
$D_s^{*+}D_s^-$&$0.84^{+0.12}_{-0.12}$&$1.2^{+0.2}_{-0.1}$&0.23&0.33&0.81&$0.92^{+0.11}_{-0.11}$&$0.70^{+0.44}_{-0.31}$&\raisebox{1.8ex}[0pt]{$2.75^{+0.83}_{-0.71}\pm0.69$}\\\hline
$D_s^{*+}D_s^{*-}$&$2.4^{+0.4}_{-0.4}$&$3.0^{+0.5}_{-0.5}$&1.48&1.3&3.2&$2.36^{+0.40}_{-0.38}$&$0.99^{+0.72}_{-0.54}$&$3.08^{+1.22+0.85}_{-1.04-0.86}$\\\hline
$D_s^+D_{s0}^{*-}$&$0.13^{+0.01}_{-0.02}$&&&&&$0.053^{+0.022}_{-0.018}$&&\\\hline
$D_s^+D_{s1}^{-}(2460)$&$0.37^{+0.12}_{-0.10}$&&&&&&&\\\hline
$D_s^+D_{s1}^{-}(2536)$&$0.039^{+0.019}_{-0.013}$&&&&&&&\\\hline
$D_s^{*+}D_{s0}^{*-}$&$0.12^{+0.02}_{-0.02}$&&&&&&&\\\hline
$D_s^{*+}D_{s1}^{-}(2460)$&$1.8^{+0.4}_{-0.4}$&&&&&&&\\\hline
$D_s^{*+}D_{s1}^{-}(2536)$&$0.21^{+0.08}_{-0.06}$&&&&&&&\\\hline
\end{tabular}\end{center} \label{tab4-3}
\end{table}

Compared our FA results and the results with corrections estimated by the pQCD approach
(we will call them the pQCD corrected results),
it can be found that the non-factorizable effects
(and annihilation contributions when exist) for $\bar{B}^0_s\rightarrow D^{(*)+}_s+L^-$
channels make little corrections to the FA results. For double charmed $B_s$ decays,
the factorizable color-favored diagrams still dominate the decay widths,
but contributions from other diagrams give up to $\sim20\%$ corrections.
We also estimate some $\bar{B}^0_s\rightarrow D^{(*)}_s+S(A)$ decay rates under the FA,
where $S$ represents scalar and $A$ stands for axial vector,
but the uncertainties due to non-factorizable effects should be realized.
The results in the 2, 4-7 columns are calculated in the FA,
but with different approaches to
evaluate form factors. For $\bar{B}^0_s\rightarrow D_s^{+}X^-$
decays, the results from the FA are roughly consistent with each other
except the values from light cone sum rules (LCSR) approach, which
are smaller than others. This discrepancy reflects the difference of
form factors, which has been shown explicitly in Table~\ref{tab4-1}.
For $\bar{B}^0_s\rightarrow D_s^{*+}X^-$ decays, our results are
consistent with HQET and CLFQM values, but larger than those in BSW
and QCDSR methods.

Now we turn to compare the our results with the pQCD results in
\cite{Li2}. The major difference between our pQCD corrected
results and their fully pQCD results is that, for calculating
factorizable color-favored contributions, we estimate them in
terms of our form factors, while in \cite{Li2} all the
contributions are calculated in the pQCD approach. Besides, some
input parameters are different between the two works. The
comparison can be seen from Table~\ref{tab4-3}: most of the pQCD
results are smaller than our FA results as well as the pQCD
corrected results, but close to LCSR results. The Isgur-Wise form
factor at maximum recoil in the pQCD is $\xi^{B_s\rightarrow
D_s}_+=0.44$, whereas the corresponding quantity in our work is
$\xi_+ = (M_{B_s}(f_+ + f_-) + M_{D_s}(f_+ -
f_-))/2\sqrt{M_{B_s}M_{D_s}}=0.55$. Besides, authors of \cite{Li2}
found that the form factors from their pQCD calculations are
similar with those from LCSR. Therefore, we can simply state that
form factors from the pQCD approach in \cite{Li2} are smaller than
ours, which means that the discrepancies in decay rates between
their pQCD results and ours could be attributed to the difference
in form factors, rather than non-factorizable effects.

Thanks to the efforts done by Belle, CDF, D0 and other
Collaborations, some experimental data of two-body non-leptonic
$B_s$ decays are available now. We can see from Table~\ref{tab4-3}
that our results agree well with the present experimental data.
Since only some $\bar{B}^0_s\rightarrow D_s^{(*)+}X^-$ decay
modes have been observed and the experimental errors are still large
till now, it is expected in the near future more precise tests could
be made on theoretical predictions as the increase of the $B_s^0$
events.

We now investigate the direct CP asymmetries of $B_s\rightarrow
D_s^{(*)}D_{q}^{(*)}$ decays. The results together with $D_1$,
$D_2$ are shown in Table~\ref{tab4-4}. $D_1$, $D_2$ are defined in
equation (\ref{eq3-13}). Their values with the weak phase $\gamma$
determine the direct CP asymmetries. We show both the CPAs
estimated under FA and in the pQCD approach. The discrepancy
between FA and pQCD (or our pQCD corrected results) is obvious and
is found mainly arising from the strong phases. It should be
mentioned that another method called QCD improved factorization,
which can cover the non-factorizable effects as well as pQCD,
makes quite different predictions on CP asymmetries of some decay
modes compared to pQCD. The reason is that the leading sources of
the strong phase are different between the two
approaches~\cite{LiHN}. According to our results in
Table~\ref{tab4-4}, most of the direct CP asymmetries are too
small to be tested experimentally for now.

\begin{table}[h] \small\caption{The direct CP asymmetries in double charmed non-leptonic decays of
$\bar{B}_s^0$ in the unit of $10^{-2}$. $D_1$ and $D_2$ are
defined as $\mathcal{A}_{CP}^{\mathrm{dir}}\equiv
D_1\frac{\sin\gamma}{1+D_2\cos\gamma}$, where $\gamma$ is the weak
phase (see equations (\ref{eq3-12},\ref{eq3-13})). In calculation,
the weak phase is taken as $\gamma=68.8^{\circ}$. We show the
results with non-factorziable and annihilation contributions
estimated in the pQCD approach. The FA results are also shown in
the brackets.} {\begin{center}\begin{tabular}{|c|cc|cc|cc|c|}
\hline Final States&\multicolumn{2}{|c|}{$D_1$ in this
work}&\multicolumn{2}{|c|}{$D_2$ in this
work}&\multicolumn{2}{|c|}{$\mathcal{A}_{CP}^{\mathrm{dir}}$ in
this work}&$\mathcal{A}_{CP}^{\mathrm{dir}}$ in \cite{Li2}
\\\hline
$D_s^+D^-$&$-0.36\sim -0.24$&(5.3)&$9.8$&(10)
&$-0.3\sim -0.2$&(4.8)&$-0.5_{-0.1-0.2}^{+0.0+0.1}$\\\hline
$D_s^+D_s^-$&$0.12\sim 0.15$&(-0.32)&-0.47&(-0.97)&$0.1$&(-0.30)&\\\hline
$D_s^+D^{*-}$&$-0.064\sim 0.039$&(1.5)&3.1&(2.6)&$-0.06\sim 0.04$&(1.4)&$-0.1_{-0.0-0.0}^{+0.1+0.0}$\\\hline
$D_s^+D_s^{*-}$&$0.048\sim 0.063$&($-0.081$)&-0.16&(-0.14)&$0.05\sim0.06$&($-0.076$)&\\\hline
$D_s^{*+}D^-$&$0.56\sim 0.71$&(-0.46)&-0.69&(-1.6)&$0.5\sim 0.7$&(-0.43)
&$0.4_{-0.0-0.1}^{+0.0+0.1}$\\\hline
$D_s^{*+}D_s^-$&$-0.074\sim-0.064$&($0.027$)&$0.032$&(0.13)&$-0.07\sim-0.06$
&($0.025$)&\\\hline
$D_s^{*+}D^{*-}$&$0.\sim0.13$&(1.5)&2.9&(2.6)&$0.\sim 0.1$&(1.4)&$-0.1_{-0.1-0.1}^{+0.0+0.0}$\\\hline
$D_s^{*+}D_s^{*-}$&$-0.0018\sim0.015$&($-0.081$)&-0.20&(-0.14)
&$-0.002\sim0.01$&($-0.076$)&\\\hline
\end{tabular}\end{center} \label{tab4-4}}
\end{table}

The results of polarization fractions of $B_s^0\rightarrow VV(A)$
decays are listed in Table~\ref{tab4-5}, compared with other
theoretical estimates and available experimental data. From the
table, we can say that the non-factorizable contributions (and
annihilation contributions when exit) do not give sufficient
corrections on factorizations. Besides, it could be found that
although our form factors are different from those in \cite{Li2},
the polariztions are still similar, which tells that the
polariztions are less affceted by the form factors. Please note
that for decays with a $D_s^*$ and a light meson in the final
state, $R_L\sim 0.8\gg R_\parallel,~R_\perp$; for decays with two
charmed mesons in the final state, $R_L\sim R_\parallel\sim0.5\gg
R_\perp$. The similar results have been found in $B\rightarrow VV$
decays.

\begin{table}[h] \small\caption{Polarization fractions of $\bar{B}_s^0\rightarrow VV$ or
$VA$ decays. $R_L$, $R_{\parallel}$ and $R_{\perp}$ are
longitudinal, transverse parallel and transverse perpendicular
polarization fractions, respectively. Our results with pQCD
represent those with non-factorziable and annihilation
contributions estimated in the pQCD approach.}
{\begin{center}\begin{tabular}{|c|c|c|c|c|c|} \hline Final
States&&Ours (FA)&Ours (pQCD)& \cite{Li2}& Experiment~\cite{Belle}
\\\hline
$D_s^{*+}\rho^-$&$R_L$&$0.874^{+0.004}_{-0.003}$&$0.854^{+0.004}_{-0.005}$&0.87&$1.05^{+0.08}_{-0.10}(\mathrm{stat})^{+0.03}_{-0.04}(\mathrm{syst})$\\\hline
&$R_{\parallel}$&$0.104^{+0.005}_{-0.004}$&$0.113^{+0.005}_{-0.005}$&&\\\hline
$D_s^{*+}K^{*-}$&$R_L$&$0.841^{+0.004}_{-0.005}$&$0.857^{+0.003}_{-0.003}$&0.83&\\\hline
&$R_{\parallel}$&$0.133^{+0.005}_{-0.006}$&$0.104^{+0.004}_{-0.004}$&&\\\hline
$D_s^{*+}a_1^-$&$R_L$&$0.738^{+0.005}_{-0.007}$&&&\\\hline
&$R_{\parallel}$&$0.221^{+0.007}_{-0.008}$&&&\\\hline
$D_s^{*+}D^{*-}$&$R_L$&$0.525^{+0.006}_{-0.006}$&$0.511^{+0.010}_{-0.006}$&$0.56\pm0.14$&\\\hline
&$R_{\parallel}$&$0.416^{+0.009}_{-0.009}$&$0.439^{+0.009}_{-0.010}$&&\\\hline
$D_s^{*+}D_s^{*-}$&$R_L$&$0.503^{+0.005}_{-0.006}$&$0.537^{+0.006}_{-0.004}$&$0.53\pm0.15$&\\\hline
&$R_{\parallel}$&$0.439^{+0.009}_{-0.010}$&$0.427^{+0.006}_{-0.012}$&&\\\hline
$D_s^{*+}D_{s1}^{-}(2460)$&$R_L$&$0.436^{+0.004}_{-0.004}$&&&\\\hline
&$R_{\parallel}$&$0.512^{+0.008}_{-0.008}$&&&\\\hline
$D_s^{*+}D_{s1}^{-}(2536)$&$R_L$&$0.424^{+0.003}_{-0.004}$&&&\\\hline
&$R_{\parallel}$&$0.527^{+0.008}_{-0.007}$&&&\\\hline
\end{tabular}\end{center} \label{tab4-5}}
\end{table}

To conclude, we note that the form factors of $B_s^0\rightarrow
D_s^{-}$ and $B_s^0\rightarrow D_s^{*-}$ estimated in this work are
roughly consistent with other theoretical estimates such as the QCD
sum rule, the light-front quark model and so on. Using the derived
form factors, branching ratios of semi-leptonic and non-leptonic
$B^0_s$ decays to charmed particles are estimated.
For non-leptonic decays we evaluate amplitudes under the factorization approximation
as well as in the pQCD approach (for non-factorizable and annihilation contributions).
Our predictions on decay rates are found to agree well with the available experimental data.
Direct CP asymmetries estimated in FA and in pQCD are quite different (even have different sign).
Generally, the CPAs of $\bar{B}_s^0\rightarrow
D_s^{(*)+}D_q^-$ are less than $1\%$. Polarization fractions of
$B_s\rightarrow D_s^*V(A)$ decays follow the similar rule as
$B\rightarrow VV$ decays, that is, for decays with a $D_s^*$ and a
light meson in the final state, $R_L\sim 0.8\gg
R_\parallel,~R_\perp$, for decays with two charmed mesons in the
final state, $R_L\sim R_\parallel\sim0.5\gg R_\perp$.

\section*{Acknowledgments}

H. Fu would like to thank Run-Hui Li for useful discussion on the pQCD approach.
The work of X.C. was supported by the Foundation of Harbin
Institute of Technology (Weihai) No. IMJQ 10000076. C.S.K. was
supported in part by the NRF grant funded by the Korea government
(MEST) (No. 2010-0028060) and (No. 2011-0017430), and in part by
KOSEF through the Joint Research Program (F01-2009-000-10031-0).
The work of G.W. was supported in part by the National Natural
Science Foundation of China (NSFC) under Grant No. 10875032, No.
11175051, and supported in part by Projects of International
Cooperation and Exchanges NSFC under Grant No. 10911140267.

\end{document}